\documentstyle[11pt,paspconf,epsf]{article}

\begin{document}
\title{The NOG Sample:\\ 
Galaxy Systems and 3D Real-Space Galaxy Density}  
\author{Christian Marinoni, Giuliano Giuricin, Lorenzo Ceriani, 
Armando Pisani} 
\affil{Dept. of Astronomy, Trieste Univ. and SISSA, Trieste, Italy}

\begin{abstract}
We discuss the  3D real-space reconstruction of the 
optical galaxy density field in the local Universe 
as derived from the galaxies of the Nearby Optical Galaxy (NOG) sample.

NOG is a distance-limited ($cz_{LG}\leq$6000 km/s) and  
magnitude--limited ($B\leq$14 mag) sample of 7076 optical  galaxies which 
covers 2/3 (8.29 sr) of the 
sky ($|b|>20^{\circ}$). We have replaced ``true distances'' 
measurements for all the objects in order to correct for redshif distortions
(Marinoni et al. 1998).

Using homogenized photometric information for the whole sample, NOG is meant 
to be an approximation to a homogeneous all-sky 3D optically 
selected and statistically well-controlled
galaxy sample that probes in great detail volumes of cosmological interest.

Our goal is to construct a reliable, robust and unbiased field of
density contrasts covering interesting regions of galaxy and mass overdensities
of the local universe.
\end{abstract}

\section{The selection of the NOG} 
We select the sample attempting to tighten the selection criteria already 
used in the pioneer work of Hudson (1993) and  Optical Redshift Survey 
(ORS, Santiago et al. 1995), specifically, 
by adopting a 
complementary approach to the construction of an all-sky optical galaxy sample.
In particular, we use, as photometric selection parameter, the  total blue 
magnitudes, homogeneously
transformed into the standard system of the RC3 catalogue and corrected 
for Galactic extinction, internal extinction and K-dimming. 
 
Though being limited to a depth of 6000 km/s, the NOG covers interesting
regions of galaxy and mass overdensities of the local universe, such as
the "Great Attractor" region and the Perseus-Pisces supercluster. 
Compared to previous all-sky optical and IRAS galaxy samples (e.g. the
ORS, the IRAS 1.2 Jy by Fisher
et al.  1995, the PSCz by Saunders et al. 1999), the NOG provides a denser
sampling of the galaxy density field in the nearby universe. NOG contains
11\% more galaxies than the ORS limited to 6000 km/s and 35\% more
galaxies than the PSCz limited to the same depth, although it covers about
3/4 of the solid angle covered by the PSCz. Besides, NOG delineates
overdensity regions with a greater contrast than IRAS samples do.

\begin{figure}
\plotfiddle{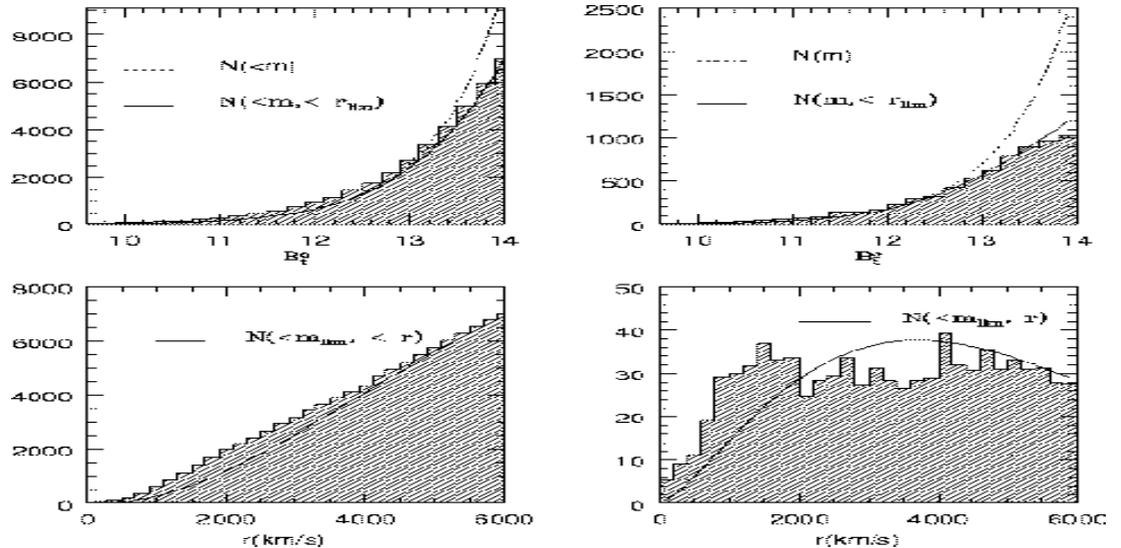}{7.5cm}{0}{85}{42.5}{-268}{-60}

\caption{Observed Integral ({\em left}) and differential ({\em right}) counts 
of NOG galaxies fitted assuming an euclidean geometry of space and a
uniform distribution of galaxies.
The solid line indicates theoretical counts predicted
inside the NOG volume ($r_{lim}$=6000 km/s).}
\end{figure}

However, the high-density sampling rate, achieved by
maximizing the sky coverage and limiting the sample depth,
is counteracted by systematic effects arising 
from the cutoffs in the selection parameters or non-uniformities in 
the original catalogs. So we have tried to correct and  minimize these
biases testing the sample completeness by means of a count-magnitude 
analysis  and deriving the appropriate  
luminosity and redshift selection functions.
We find that the NOG is intrinsically complete
down to its limiting magnitude B=14 mag.
Moreover, it has a good 
completeness in redshift ($\sim$98\%) (see fig. 1).

We have evaluated the Schechter luminosity
function for the whole sample and for different morphological 
types (see Marinoni et al. 1999 for preliminary results)

\section{Identification of galaxy systems}

Since we are interested in describing the galaxy density field also on
small physical scales, we identify galaxy systems in the NOG in order to remove
non linearities on small scales in the peculiar velocity field and also
study clustering properties and statistics.

This has been done (Giuricin et al. 1999) by means of the
hierarchical method and percolation {\it friends of friends} method.
We obtain two homogeneous catalogs of loose groups which turn out to be
substantially consistent. Most of the NOG galaxies ($\sim$60\%) are found to be
members of galaxy pairs ($\sim$580 for a total of $\sim$15\% of the
galaxies) or groups with at least three members ($\sim$500 groups
comprising $\sim$45\% of the galaxies). About 40\% of the galaxies are
left ungrouped (field galaxies). Containing about 500 groups with at least 
three members,  they are among the largest catalogs of groups presently 
available in the literature.

Furthermore, we have evaluated the Schechter luminosity
function for field galaxies, group members, 
and for different morphological types (see Marinoni et al. 1999 for 
preliminary results).

\begin{figure}
\plottwo{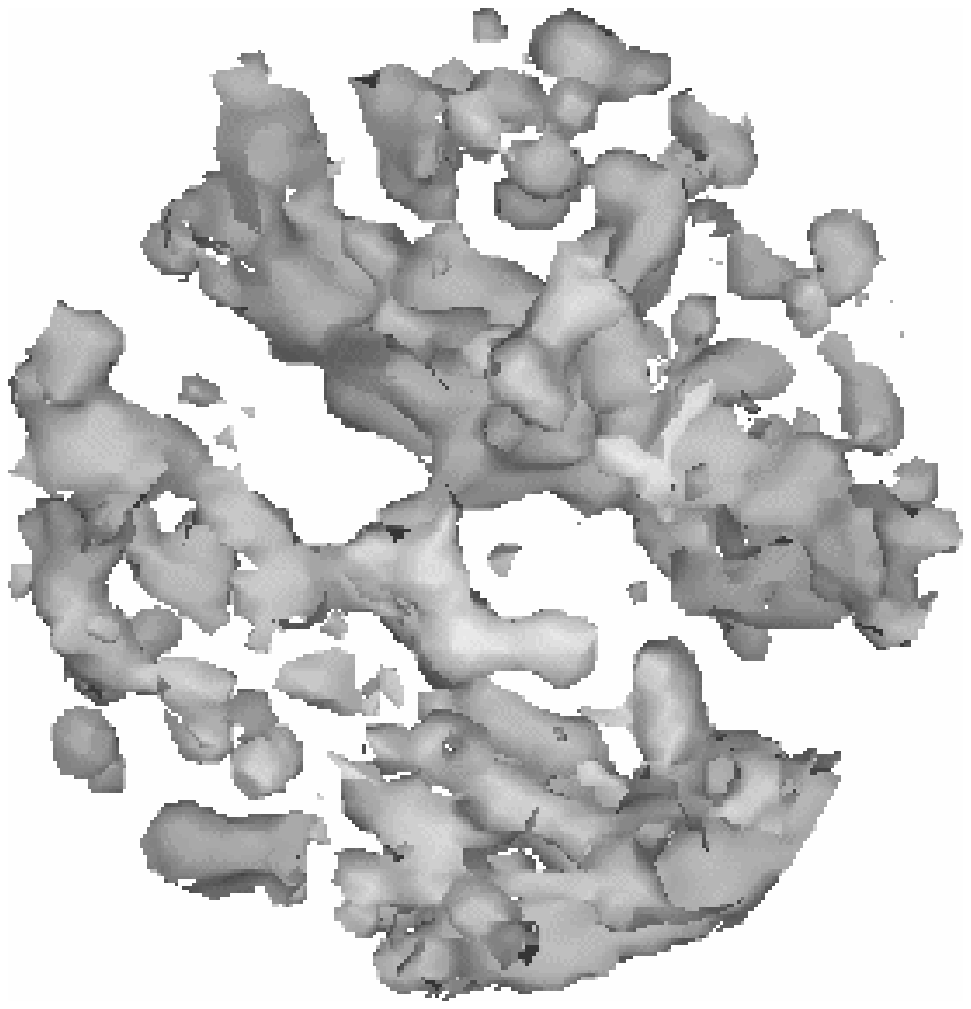}{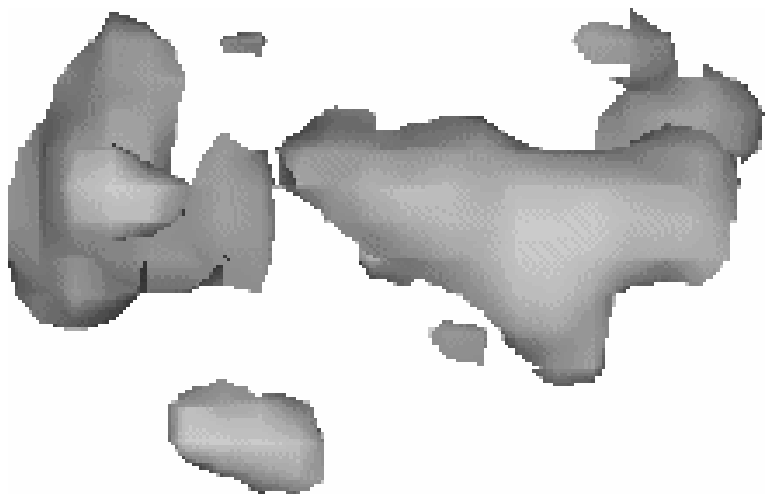}
\caption{3D galaxy distribution in the nearby universe:
surface contours of density contrast $\delta=1.5$ are plotted using 
a small, $\langle r_s \rangle =200 
km/s$, ({\em left})
and a large,  $\langle r_s \rangle =500 km/s$, ({\em right}) 
smoothing length.
The plot on the right reveals the flattened structure of Perseus Pisces 
and the continuity between the Local Supercluster and the Great Attractor
complex}
\end{figure}

\section{Density Reconstruction}

The problem of reconstructing the density fluctuation $\delta({\bf r})$
is connected with finding the best transformation scheme for {\em diluting}
the point distribution into a continuous density field reducing the
flux-limited effects and shot noise problems. 

We address all these issues smoothing with a normalized Gaussian filter
characterized by a smoothing length which is a properly defined 
increasing function of distance. Moreover, we assign each galaxy a weight 
given by the inverse of the sample selection function in order to well
calibrate the median value of the density (see Marinoni et al. 1999).

The specific features of the galaxy distribution field
are shown in figure 2 where we plot the surface contours of 
the 3D galaxy density contrast $\delta=1.5$ using 
a small and a large smoothing length (respectively $\langle r_s \rangle =200 
\; km/s$ and  $\langle r_s \rangle =500 \; km/s$). 

It is clear how NOG can constrain the shape and  dimensionality of 
high-amplitude, nearby structures as the so-called Supergalactic Plane. 
One's first visual impression is the irregularity in the 
the shapes of the major structures and the first-order symmetry 
between high- and low-density regions.

\section{Conclusion}

Given its large sky coverage, its high-density sampling, the 
identification of galaxy systems and real-space distance information,
the NOG is well suited for  mapping the 
cosmography of the nearby universe, studying the clustering properties 
and tracing the optical galaxy density field (also on small scales) to 
be compared with the IRAS galaxy density field.

Exploring in detail the nature of the 3D galaxy distribution
will provide  us with invaluable qualitative cosmographical
information about the  topology and morphology of the local 
overdensities
but also will allow us to investigate on the  z=0 cosmology,  
greatly increasing our  quantitative understanding of
physical parameters that  constrain the evolution of  
structures and their clustering properties.

An important issue in the field of observational cosmology is the possible 
existence of biasing in the galaxy distribution relatively to the matter. 
It is reasonable to expect that besides  depending on position and
scale, the amount of bias should be connected to some intrinsic galaxian 
property. Thus, the comparison of the relative distribution or morphological
segregation of different types of galaxies, for which NOG provide a detailed 
classification, will help us in constraining different theories for structure 
formation and evolution.

Moreover its nearly full-sky nature and the large variety in galaxy
content  make the NOG ideal also for more specific tasks as the  
deconvolution of environmental effects from the properties and evolution
history of the galaxies.

\bigskip

\acknowledgments We wish to thank S. Borgani, D. Fadda,
M. Girardi, M. Hudson, F. Mardirossian,  M. Mezzetti, P. Monaco, 
M. Ramella, and S. Samurovi\'c for interesting conversations.

\end{document}